\documentstyle[12pt]{article}
\begin{document}
\begin{center}
{\large\bf SOLAR NUCLEAR ENERGY GENERATION AND THE CHLORINE SOLAR
NEUTRINO EXPERIMENT}\\ [0,5cm]
{\bf H.J. HAUBOLD}\\
UN Outer Space Office, Vienna International Centre, Vienna,
Austria\\[0,5cm]
and\\ [0,5cm]
{\bf A.M MATHAI}\\
Department of Mathematics and Statistics, McGill University,
Montreal, Canada\\[1cm]
{\bf ABSTRACT}
\end{center}
\bigskip
The study of solar neutrinos may provide important insights into
the physics of the central region of the Sun. Four solar neutrino
experiments have confirmed the solar neutrino problem but do not
clearly indicate whether solar physics, nuclear physics, or
neutrino physics have to be improved to solve it. Nonlinear
relations among the different neutrino fluxes are imposed by two
coupled systems of differential equations governing the internal
structure and time evolution of the Sun. We assume that the results
of the four neutrino experiments are correct and are concerned not
with the discrepancy between the average rate and the predicted
rate, but with a possible time dependence of the argon production
rate as revealed in the Homestake experiment over a time period of
20 years. Based on the subtlety of the solar neutrino problem we
review here qualitatively the physical laws employed for
understanding the internal solar structure and conjecture that the
interlink between specific nuclear reactions of the PPIII-branch
of the proton-proton chain may allow the high-energy solar neutrino
flux to vary over time.
\clearpage
\section{Solar Neutrino Detection: Results}
The Sun is supposed to be a simple main-sequence star that
generates its energy through the proton-proton chain and to a much
less extend through the CNO cycle, thereby producing a copious flux
of neutrinos. The Sun is considered to be a simple star because very
basic physical laws can be employed to describe the evolution and
the internal structure of the Sun, namely Newton's laws of gravity
and motion, the first two laws of thermodynamics, Einstein's law of
equivalence of mass and energy, Boyle's law and Charles' law of
perfect gases, and Heisenberg's uncertainty principle. To probe the
simplicity and consistency of theoretically developed standard
solar models, over the past two and a half decades four experiments
to detect the solar neutrino flux have been established (Bahcall
and Pinsonneault 1992).\\
(1) The chlorine experiment in the Homestake gold mine (USA), in
operation since 1967, observed an average $^{37}$Ar production rate
by solar neutrinos with energies $E_\nu >0.8$MeV of $(2.2\pm
0.2)$SNU\footnote{1SNU (solar neutrino unit)=$10^{-36}$ captures
per atom per second} for runs 18-109 over the time period 1970.8 to
1990.0 (Davis 1993).\\
(2) The neutrino-electron scattering experiment in the Kamiokande
mine (Japan), in operation since 1986 for two time periods of 1040
days (Kamiokan-
de II: Jan87-Apr90, 450 days with $E_\nu>9.3$ MeV and
590 days with $E_\nu >7.5$ MeV) and 220 days (Kamiokande III:
Dec90-Dec91, $E_\nu >7.5$ MeV), detected solar neutrinos with event
rates $<\Phi (^8B)_{obs} >=[0.47\pm0.05(stat)\pm0.06
(syst)]\Phi_{calc}(^8B)$ of that predicted by standard solar models
(Nakamura 1993).\par
For the chlorine experiment the standard solar model with the best
input parameters predicts an event rate of $(8.0\pm3.0)$ SNU
(Bahcall and Pinsonneault 1992). The experimental results referred
to in (1) and (2) above revealed that the measured solar neutrino
fluxes are significantly below those predicted by standard solar
models, known as the ``solar neutrino problem''.\\
(3) The gallium experiment under a mountain in the North Caucasus
at Baksan (Russia) began to detect solar neutrinos with energies
$E_\nu>0.23$ MeV in 1988 and reported a $^{71}Ge$ production rate
of $(85^{+22}_{-32}[stat]\pm 20[syst])$ SNU (Anosov et al. 1993).\\
(4) The European gallium experiment in the Gran Sasso tunnel
(Italy), in operation since 1991, measured a capture rate of solar
neutrinos by $^{71}Ga$ of $(79\pm 10[stat]\pm 6[syst])$ SNU in 30
runs
between May 1991 (GALLEX I: May91-March92) and October 1993
(GALLEX II: Aug92-Oct93)(Anselmann et al. 1994).\par
The detected solar neutrino fluxes for the two gallium experiments
almost agree with each other but are in conflict with standard
solar models which predict a capture rate of $(132^{+21}_{-17})$SNU
(Bahcall and Pinsonneault 1992).\par
The result from the neutrino-electron scattering experiment
(Kamiokande) constrains only the high-energy $^8B$ neutrino flux,
while the chlorine and gallium experiments (Homestake, Baksan, Gran
Sasso) constrain both the $^7Be$ and $^8B$ solar neutrino fluxes.
The results of the four neutrino experiments, despite of extensive
experimental and theoretical efforts to reveal the origin of the
discrepancies, do not indicate clearly whether solar physics,
nuclear physics, or neutrino physics have to be improved to settle
the account of the solar neutrino problem.\par
The net reaction for the proton-proton chain and the CNO cycle is
the conversion of hydrogen into helium,
\[4p\longrightarrow \alpha + 2\nu_e+2e^+ +Q, Q=26.7 MeV,\]
where two neutrinos are produced in the Sun per 26.7 MeV release
of nuclear energy. This reaction allows an estimate of the copious
solar neutrino flux on Earth, assuming that the solar nuclear
energy generation equals the Sun's luminosity and does not vary
over time periods short in comparison to the nuclear time scale,
$$\Phi_{\nu\odot}=\frac{2L_\odot}{Q-
2E_\nu}\frac{1}{4\pi(AU)^2}\approx 6.5\times 10^{10}\nu_\odot
cm^{-2}s^{-1},$$
where $L_\odot=3.86\times 10^{33}ergs^{-1}$ is the Suns's
luminosity, $AU=1.5\times10^{13}cm$ is its average distance from
Earth, and $E_\nu\approx 0.26$ MeV is the average energy of the
produced neutrinos. The net reaction assumes that baryon number,
charge flavour, and energy are conserved quantities. {\it In the
following it will be assumed that the reported results of the four
neutrino experiments are correct.}
\section{Fundamental Physical Constants and
Solar Structure: Differential Equations}
There are two distinct types of basic equations in physics:
dynamical equations exhibiting the reversible (Newtonian) time and
transport equations reflecting the irreversible
(Boltzmann-Gibbsian) time. The Sun contains a large number of
hydrogen atoms,
and evolves as a main-sequence star in nuclear time scale, an
extremely long time in comparison to thermal diffusion time
(Helmholtz-Kelvin time scale) and to its fundamental pulsation
mode. Accordingly, the time evolution of the Sun is managed by the
change of its chemical composition governed by a system of coupled
nonlinear kinetic equations which can be solved separately from the
system of coupled partial differential equations of solar
structure, boundary conditions, and the constraint that the model
luminosity at the present epoch must be equal to the observed solar
luminosity.\par
The four basic differential equations required to calculate the
internal structure of the Sun are equations which represent
respectively the distribution of mass within the Sun, the balance
of gravity and pressure giving hydrostatic equilibrium (involving
Newton's gravitational constant G), the outward flow of energy
driven by the temperature gradient inside the Sun (involving the
velocity of light c and Stefan's constant a), and the equation of
nuclear energy generation within the Sun which continually
replenishes that radiated away (involving Planck's quantum of
action $\hbar$). In order to solve these differential equations, it
is necessary to specify the equation of state (involving
Boltzmann's constant k), the opacity of the solar material and the
nuclear energy generation rates. Boundary conditions have to be
satisfied at the surface and at the centre of the Sun. From a more
general point of view with regard to the calculation of the
internal structure of the Sun it is only necessary to make an
assumption about the distribution of the energy sources within the
Sun, not necessarily knowing what produces the energy. Due to this
fact, calculations of the internal structure of solar-type stars
made already considerable progress before the production of energy
by nuclear synthesis was understood in any detail (Eddington 1928,
cp. Chandrasekhar 1984). {\it The differential equations for the
internal structure of the Sun contain fundamental constants of
physics that make it possible to gain qualitative insight into the
solutions of the system of differential equations by simple
dimensional analysis.} In quantum field theory the Sommerfeld fine-
structure constant $\alpha_{el}$ plays the role of a dimensionless
coupling constant for the Coulomb force:
$$\alpha_{el}^{-1}=\frac{\hbar c}{e^2}\approx 137.$$
Equivalent to this quantity one can define the gravitational
fine-structure constant $\alpha_g$,
$$\alpha_g^{-1}=\frac{\hbar c}{Gm_p^2}\approx 10^{38},$$
where $m_p$ is the mass of the proton. This quantity measures the
smallness of the gravitational force between two protons, similar
to $\alpha_{el}$ which measures the smallness of the Coulomb force
between two electrons. In terms of stability of matter,
$\alpha_{el}^{-1}$ gives the maximum positive charge of the central
nucleus that will allow a stable electron-orbit around it. It can
be further shown that the combination of the fundamental
constants in $\alpha_{el}^{-1}$ and $\alpha_g^{-1}$ are playing an
important role in the
evolution and structure of the Sun as expected from the
differential equations (Salpeter 1966). Electromagnetic
and
gravitational interactions are long range interactions, where the
unit of electromagnetic interaction is e and the unit of
gravitational interaction is $m_p  (m_p=1837 m_e).$ The fourth
fundamental constant of physics, Boltzmann's k, is an exceptional
case since temperature can be defined in terms of energy (Gamow
1970). However,
if the equation of state for solar material followed exclusively
the perfect gas law, one would not have any preferred units for
density and temperature. This leads necessarily to the
consideration of radiation pressure (involving Planck's constant
$\hbar$) and electron degeneracy (involving $\hbar$ and the
electron
mass $m_e$) for the internal structure of the Sun.
\section{The Sun is Massive Because Gravity is Weak: Virial
Theorem}
The Sun is a system of N nucleons with a mean separation of the
order of magnitude d,
$$M_\odot \approx N m_p,\; R_\odot \approx N^{1/3}d,$$
where $M_\odot$ and $R_\odot$ denote the mass and radius of the
Sun, respectively. Observationally the Sun ought to be considered
close to hydrostatic equilibrium while from the theoretical point
of view the Sun is considered to be in complete hydrostatic
equilibrium. The virial theorem implies that the gravitational
binding energy of the Sun must be of the order of its internal
energy,
$$3\int dVP = \frac{GM^2_\odot}{R_\odot}=-\Omega_g=\frac{Gm_p^2
N^{5/3}}{d},$$
where P is the total pressure in volume element dV and the
right-hand side of the equation is the total gravitational
potential
energy of the Sun. The total thermal energy content of the Sun is
the kinetic energy per particle times its number,
$$E_{th} \simeq NkT.$$
Two forces balance to keep the Sun in hydrostatic equilibrium: the
gravitational force directed inward and the gas and radiation
pressure force directed outward. The total radiation energy is the
product of the volume of the Sun and $aT^4:$
$$E_r\simeq Nd^3aT^4,$$
where a is Stefan's constant,
$a=\frac{\pi^2}{15}\frac{k^4}{c^3\hbar^3}$, by virtue of Planck's
law. According to Heisenberg's uncertainty principle, the Fermi
momentum of free electrons is $p_F \approx \hbar /d,$ where d is
the average separation of electrons. Provided the electrons are
nonrelativistic, their Fermi energy is $E_F=p^2/2m_e$ and one
obtains for the electron degeneracy energy,
$$E_d \simeq N\frac{p^2}{2m_e}=N \frac{\hbar^2}{2m_e d^2}.$$
Thus, the virial theorem implies that
$$NkT+Nd^3aT^4+N\frac{\hbar^2}{2m_e d^2}\approx
N^{5/3}\frac{Gm_p^2}{d}.$$
For a star of solar mass with a mean molecular weight equal to 1,
the radiation pressure at the center cannot exceed a few percent of
the total
pressure and can be neglected. Similarly, electron degeneracy does
not contribute to the total pressure under solar conditions. Hence,
$$kT \simeq N^{2/3} \frac{Gm_p^2}{d}=N^{2/3}\alpha_g
\frac{\hbar c}{d}=\left(\frac{N}{N_0}\right)^{2/3}\frac{\hbar c}{
d},$$
with $N_0=\alpha_g^{-3/2}.$ {\it The Sun is massive because gravity
is
weak.} This simple relationship between temperature and number of
nucleons also confirms that the most important fact concerning a
star is its mass. More detailed calculation shows that there exists
an upper limit for the mass of a stable star of $\sim 100 N_0 m_p,$
otherwise radiation would dominate and lead to the disruption of
it. Likewise a lower limit for a star's mass can be derived, $\sim
0.1 N_0 m_p,$ to account for the temperature needed to ignite
nuclear fuel to form a self-supported shining star, i.e., to burn
hydrogen (Weisskopf 1975, Carr and Rees 1979, Dyson 1979).\par
The internal structure of a configuration $M\simeq 100 N_0 m_p$ is
dominated by radiation pressure, while for $M\geq 0.1 N_0 m_p$ it
is held together by electron degeneracy.
\section{Gravitationally Stabilized Solar Fusion Reactor:
Adjustment Factor}
The basic condition for thermonuclear reactions between charged
particles is that their thermal energy must be large enough to
penetrate the Coulomb repulsion between them. Nuclear reactions are
collision phenomena characterized by cross sections. The cross
section $\sigma$ of a reaction is defined as the probability that
the reaction will occur if the incident flux consists of one
particle and the target contains only one nucleus per unit area.
The microscopic nature of the particles requires the quantum
mechanical treatment of the collision problem. The number of
reactions is directly proportional to the number density of the
incident flux and the number density of the target. In the case of
the nuclear fusion plasma within the Sun, thermal equilibrium is
commonly assumed for the ensemble of nuclei. The distribution of
the relative velocities among the nuclei is Maxwell-Boltzmannian.
The thermonuclear reaction rate is given by
$$r_{12}=n_1n_2<\sigma v>_{12},$$
where $n_1$ and $n_2$ denote the number densities of particles of
type 1 and 2, respectively, and $<\sigma v>_{12}$ is the reaction
probability in the unit volume per unit time. This definition of
the reaction rate reveals immediately that the quantity
$$\tau_{12}=[n_2<\sigma v>_{12}]^{-1},$$
has the dimension of time and can be considered to be the
lifetime of particle 2 against reaction with particle 1. A
suitable representation of the nuclear cross section contains two
factors: A geometrical factor to which quantum mechanical
interaction between two particles is always proportional,
$\lambda^2 \sim (\mu v^2)^{-1}$ (where $\lambda$ is the
reduced de Broglie wave length, and $\mu$ is the reduced mass) and
the probability for two particles of charge $Z_1e$ and $Z_2e$ to
penetrate their electrostatic repulsion:
$$\sigma(v)=\frac{2S}{\mu v^2}exp\left\{-2\pi\frac{Z_1Z_2e^2}{\hbar
v}\right\}.$$
The constant S is called astrophysical cross section factor and
absorbs the intrinsical nuclear parts of the probability for the
occurrence of a nuclear reaction. Then, the reaction probability is
defined as the product of the cross section $\sigma$ and the
relative velocity $v$, averaged over the Maxwell-Boltzmann
distribution of relative velocities of the reacting particles,
$$f(v)dv=\left(\frac{\mu}{2kT}\right)^{3/2}exp\left\{-\frac{\mu
v^2}{2kT}\right\}4\pi v^2dv.$$
To investigate the competition between the exponential factors
contained in the Maxwell-Boltzmann distribution function and the
Gamov penetration factor the following order of magnitude
estimation is pursued. For the number density of the particle gas
we use the mean density of the Sun with mass $M_\odot$ and radius
$R_\odot$ normalized to the mass of the proton, $m_p,$
$$n_2=\frac{M_\odot}{R^3_\odot}\frac{1}{m_p}.$$
The velocity of the nuclei is assumed to be the root-mean-square
velocity of the Maxwell-Boltzmann distribution,
$$v_{12}=\left(\frac{4kT}{m_p}\right)^{1/2}.$$
The nuclear energy generated in the Sun, which is lost by
radiation, can be estimated in writing
$$E_{nuc}\approx X \Delta m M_\odot c^2,$$
where X is the fraction of mass the Sun can use for nuclear energy
generation, $\Delta m M_\odot c^2$ is the fraction of mass of the
Sun really converted into radiation energy. Thus, the nuclear
lifetime of the Sun is of the order
$$\tau^{-1}\approx \frac{L_\odot}{E_{nuc}} \approx \frac{L_\odot}{X
\Delta m M_\odot c^2}.$$
For the lifetime of particle 2 one has
$$\frac{1}{\tau_{12}}\approx \frac{L_\odot}{E_{nuc}}\approx
n_2\sigma_{12}v_{12}.$$
Thus,
$$\frac{L_\odot}{X \Delta m M_\odot c^2}\approx
\frac{M_\odot}{R_\odot^3 m_p}\frac{2S}{m_p^{1/2}(kT)^{1/2}}
exp\left\{-\frac{2\pi
e^2}{\hbar}\left(\frac{m_p}{4kT}\right)^{1/2}\right\},$$
and isolating the exponential term in this expression by setting it
equal to unity and than taking the logarithm, one gets
$$2\pi \alpha_{el}\left(\frac{m_p c^2}{4kT}\right)^{1/2}\approx
ln\left\{\frac{2M_\odot ^2 X \Delta m Sc^2}{L_\odot R_\odot^3
m_p^{3/2}(kT)^{1/2}}\right\}.$$
The numerical value of the logarithmic term on the right-hand-side
in this equation is relatively insensitive to the values inserted
for
the various quantities in the brackets. Using solar values for the
quantities, $M_\odot \approx 2\times 10^{33}g, L_\odot \approx
3.86\times 10^{33} erg s^{-1}, R_\odot \approx 7\times 10^{10}cm,
T_{c\odot} \approx 10^7 K, X=0.1, \Delta m =0.007, S_{pp}= 4\times
10^{-22}$ keV barn, one obtains for the logarithmic term a
numerical value of about 10. Then one obtains
$$kT=(\frac{(2\pi \alpha_{el})^2}{2^210^2})m_pc^2 \approx 5 keV.$$
This is the central temperature of the stationarily thermonuclear
burning Sun. Actual central temperatures are about a factor 5
smaller or larger than this value due to the important fact that
the majority
of nuclear reactions occures in the high-energy tail of the
Maxwell-Boltzmann distribution function. {\it The Sun has to
adjust this
temperature through the competition between the distribution
function of relative energies of the particles and the penetration
factor of the reacting particles.}\par
\section{Numerical Simulation of the Five-Point Moving Average of
the Argon-Production Rate of the Chlorine Solar Neutrino Experiment
: Variations Over Time}
The solar neutrino flux has been inferred from the neutrino capture
rate in the chlorine neutrino experiment, measured over the past
two
decades in 86 separate runs (1970.8-1990.0). The chlorine detector
with its threshold of 0.8 MeV is sensitive to the high energy, low
flux part of the neutrino spectrum of the Sun, principally the
neutrinos from the $^7Be$ and $^8B$ reactions. In the
following we are not concerned with the discrepancy between the
average rate and the predicted rate, but with a possible time
dependence of the measured argon-production rate.\par
\vspace{8cm}
\noindent
{\footnotesize Fig.1. Shows the relative importance of the
different
branches of
the proton-proton chain: PPI, PPII, PPIII, producing respectively
26.23 MeV, 25.62 MeV, and 19.29 MeV. The differences in the energy
production are due to the energy loss carried off by the neutrinos.
The relative importance of the different branches depends on the
nuclear reaction rates and on the temperature and density structure
inside the Sun. The lifetime of the first particle
in each reaction is indicated.}\par
The proton-proton chain begins by fusion of two protons (Figure 1).
This reaction produces the great majority of solar neutrinos
$(\Phi_\nu(pp)\approx 10^{10}\nu cm^{-2}s^{-1}, \Phi_\nu(pp)\sim
T_c^{-1.2}),$ in which a proton decays into a neutron in the
immediate vicinity of another proton; the two particles form a
heavy variety of hydrogen known as deuterium, along with a positron
and a neutrino. The deuterium nucleus produced by the pp-reaction
fuses with another proton to form $^3He$ and a gamma ray. Most
often, 86\% of the time, the PPI-branch is completed when two
$^3He$
nuclei fuse to form an alpha particle and two protons, which
return to the beginning of the cycle. Approximately 14\% of the
time, however, $^3He$ instead fuses with an alpha particle,
producing $^7Be$ and a gamma ray; the $^7Be$ then
captures an electron, transmutes into $^7Li$ and emits a
neutrino $(\Phi_\nu(^7Be)\approx 10^10 \nu cm^{-2}s^{-1}, \Phi_\nu
(^7Be) \sim T^8_c).$ $^7Be$ fuses with a proton to produce two
alpha particles and thus terminates the PPII-branch. In rare cases,
about 0.02\% of the time, $^7Be$ fuses with a proton to
produce radioactive $^8B$, which beta decays into unstable
$^8Be$ and ultimately decays into two alpha particles, a
positron and an energetic neutrino $(\Phi_\nu(^8B)\approx 10^7 \nu
cm^{-2}s^{-1}, \Phi_\nu(^8B) \sim T_c^{18}).$ This reaction
terminates the PPIII-branch and is extremely sensitive to the
actual
central temperature of the Sun.\par
\vspace{7,5cm}
{\footnotesize Fig.2. Spatial distribution of neutrino sources in
the gravitationally stabilized solar fusion reactor. The hep
reaction occurs only rarely and does not influence the rates of the
other reactions.}\par
\clearpage
Because the beta decay of $^8B$ follows the fusion of $^7Be$
and a proton, all nonstandard solar models predict more reduction
of the $^8B$ neutrino flux than the $^7Be$ neutrino flux.
Any reduction of the $^7Be$ production rate affects ultimately
both the $^8B$ and $^7Be$ neutrino flux equally. This fact
seems to make cooler Sun models incompatible with the experimental
data as the higher Kamiokande observed rate relative to the
Homestake rate cannot be explained because cooler Sun models reduce
the expected $^8B$ flux more than the $^7Be$ flux. Unless
there is an {\it independent mechanism} to suppress only the $^7Be$
neutrino emission, all nonstandard solar models are in
contradiction to the solar neutrino data collected in the
Kamiokande and Homestake experiments. The only astrophysical
explanation of this phenomenon would be a greater reduction of
temperature in the region of the energy production (i.e. in the
vicinity of $0.1 R_\odot$) than just the central part where $^8B$
neutrinos are produced (cp. Figure 2).\par
\vspace{10cm}
{\footnotesize Fig.3. Five-point moving average of the
argon-production rate versus time (in years)
in the chlorine solar neutrino experiment for
86 runs in the time period 1970.8 to 1990.0.}\par
\clearpage
Fig.3 shows a five-point moving average of the argon-production
rate, removing high frequency noise from the actual time series
collected in the chlorine solar neutrino experiment (Haubold and
Beer 1992). One notes in the five-point moving average that in the
periods 1977 to 1980 and 1987 to 1990 a suppression of the
argon-production rate occurs. The overall shape of the five-point
moving
average suggests that there are two distinctive epochs spanning the
time periods 1970 to 1980 and 1980 to 1990, respectively. Each
epoch shows a shock-like rise and after that a rapid decline of the
argon-production rate. {\it The following simple mathematical model
is
able to simulate just this behaviour of the solar neutrino flux
based on the lifetimes of concerned nuclear reactions of the
PPIII-branch}. For
the time dependence of each  of the relevant individual reactions
in the PPIII-branch, the production rates for argon are assumed to
follow
$$y=kt,$$
where $y$ denotes the argon production rate and $t$ is the time.
Then it is assumed that after a certain period the argon-production
rate decreases to zero thus forming a triangle. This formulation is
repeated over time. This type of simulation is motivated by a
growth-decay mechanism discussed in Mathai (1993). Instead of a
linear growth and decay a nonlinear growth and decay resulting in
a bell-shaped function is also considered. That is,
$$y=zr\sum^v_{j=0}exp\left\{-\frac{(t-
[r+x+2jr])^2}{2(b)^2}\right\},$$
where $z=k=\sqrt{3}$ is the slope of an equilateral triangle,
$b=\sigma$ the confidence level, $x$ denotes one set of triangles
delayed by $x$ units, $v$ is the number of respective triangles,
and $r$
is the half of the triangle base. The two epochs in Fig.3 span a
time period of approximately 10 years each which will be divided
into 40 time units of 2.5 months each. The following table contains
the parameters chosen for the three relevant nuclear reactions in
the PPIII-branch:
\begin{table}[tbh]
\vspace{0.2in}
\begin{tabular}{|l|c|l|c|}\hline
nuclear reaction & lifetime & number of triangles/ &
confidence level \\
of PPIII-branch &  & half of the triangle & \\
& & base & \\ \hline
$^7Be(e^-, \nu)^7Li$ & $10^{-1}$yr & 40 months / 20 & 6.7\\ \hline
$^7Li(p, \alpha)^4He$ & $10^{-5}$yr & 8 month / 4 & 1.3\\ \hline
$^8B(e^+,\nu)^8 Be^*$ & $10^{-8}$yr & 5 month / 2.5 & 0.8 \\ \hline
\end{tabular}
\caption{}
\end{table}
\medskip
\noindent
For the simulation of the shape of the two epochs in Fig. 3 based
on the parameters for the three nuclear reactions of PPIII-branch
contained in Table 1 one has,
\begin{eqnarray*}
y & = & z\left\{q\sum^u_{i=0}exp\left\{-\frac{(t-
[q+2iq])^2}{2(a)^2}\right\}+\right.\\
& & r\sum^v_{j=0}exp\left\{-\frac{(t-
[r+x+2jr])^2}{2(b)^2}\right\}+\\
& &\left. s\sum^w_{k=0}exp\left\{-\frac{(t-
[s+2ks])^2}{2(c)^2}\right\}\right\}.
\end{eqnarray*}
The result of the numerical computation for\\
$y(t;a,b,c;z;q,r,s;x;u,v,w)=y(t;0.8,1.3,6.7;1.7;2.5,4,20;3;7;4,0)$
is shown in Fig. 4.\par
\vspace{9cm}
{\footnotesize Fig.4. Numerical simulation of one 10-year epoch
revealed
by the
time dependence of the five-point moving average of the argon
production rate. The simulation is based upon the lifetimes of
nuclear reactions  of the PPIII-branch, producing the $^7Be$
and $^8B$ neutrinos.}
\begin{center}
{\bf REFERENCES}
\end{center}
O.L. Anosov et al., Nucl. Phys. Suppl. \underline{B31}, 111
(1993).\par
\medskip
\noindent
P. Anselmann et al., Phys. Rev. Lett., submitted
(1994).\par
\medskip
\noindent
J.N. Bahcall and M. Pinsonneault, Rev. Mod. Phys. \underline{64},
885 (1992).\par
\medskip
\noindent
B.J. Carr and M.J. Rees, Nature \underline{278}, 605 (1979).\par
\medskip
\noindent
S. Chandrasekhar, Rev. Mod. Phys. \underline{56}, 137 (1984).\par
\medskip
\noindent
R. Davis Jr., in Frontiers of Neutrino Astrophysics (Eds. Y.
Suzuki\par
and K. Nakamura), Universal Academy Press, Inc., Tokyo,
1993, pp. 47.\par
\medskip
\noindent
F.J. Dyson, Rev. Mod. Phys. \underline{51}, 447 (1979).\par
\medskip
\noindent
G. Gamow, The Three Kings of Physics. In Physics, Logic, and
History\par
(Eds. W. Yourgrau and A.D. Breck), Plenum Press, New York\par
1970, pp.203.\par
\medskip
\noindent
H.J. Haubold and J. Beer, Solar Activity Cycles Revealed by Time
Series\par
Analysis of Argon-37, Sunspot-Number, and Beryllium-10 Records.
In\par
Solar-Terrestrial Variability and Global Change (Eds. W.
Schroeder\par
and J.P. Legrand), Proceedings of the IUGG/IAGA General
Assembly,\par
Vienna 1991, pp.11.\par
\medskip
\noindent
A.M. Mathai, Can. J. Statistics \underline{21}, 277 (1993).\par
\medskip
\noindent
K. Nakamura, Nucl. Phys. Suppl. \underline{B31}, 105 (1993).\par
\medskip
\noindent
E.E. Salpeter, Dimensionless Ratios and Stellar Structure.\par
In Perspectives in Modern Physics (Ed. R. Marshak), John Wiley
and\par
Sons, New York, 1966, pp. 463.\par
\medskip
\noindent
V.F. Weisskopf, Science \underline{187}, 605 (1975).
\end{document}